\documentclass[1p,11pt]{elsarticle}
\usepackage{amsmath,amsfonts,amssymb,epsfig,color,sublabel,epstopdf}
\biboptions{compress}

\newcommand{\Un}[2]{\ensuremath{\mathrm{U}^{ #1 }( #2 )}}

\newcommand{\ket}[1]{\ensuremath{\left| #1 \right\rangle}}

\newcommand{\ME}[3]{\ensuremath{\langle #1 | #2 | #3 \rangle}}

\newcommand{\etal}{\emph{et al.}}

\newcommand{\half}{\ensuremath{\textstyle{\frac{1}{2}}}}

\newcommand{\dimN}{\ensuremath{{\mathcal N}}}

\begin{document}

\begin{frontmatter}

\title{Program in C for studying characteristic properties of two-body interactions \\ in the framework of spectral distribution theory
}
\author{K. D. Launey}
\author{S. Sarbadhicary, T. Dytrych, and J. P. Draayer}
\address{Department of Physics and Astronomy, Louisiana State University,\\
Baton Rouge, Louisiana 70803, USA
}
\date{\today}

\begin{abstract}
We present a program in C that employs spectral distribution theory for studies of characteristic properties of a many-particle quantum-mechanical system and the underlying few-body interaction. In particular, the program focuses on two-body nuclear interactions given in a $JT$-coupled harmonic oscillator basis and calculates correlation coefficients, a measure of similarity of any two interactions, as well as Hilbert-Schmidt norms specifying interaction strengths. An important feature of the program is its ability to identify the monopole part (centroid) of a 2-body interaction, as well as its `density-dependent' one-body and two-body part, thereby providing key information on the evolution of shell gaps and binding energies for larger nuclear systems. As additional features, we provide statistical measures for `density-dependent' interactions, as well as a mechanism to express an interaction in terms of two other interactions. This, in turn, allows one to identify, e.g.,  established features of the nuclear interaction (such as pairing correlations) within a  general Hamiltonian. The program handles the radial degeneracy for  `density-dependent'  one-body interactions and together with an efficient linked list data structure, facilitates studies of nuclear interactions in large model spaces that go beyond valence-shell applications. \newline

\newline

\noindent
{\bf PROGRAM SUMMARY}\newline \newline
{\it Program title:} {\tt sdt} \newline
{\it Catalogue identifier: } AEQG\_v1\_0 \newline
{\it Program summary URL}: {\tt http://cpc.cs.qub.ac.uk/summaries/AEQG\_v1\_0.html} \newline
{\it Program obtainable from}: CPC Program Library, QueenÕs University, \newline  Belfast, N. Ireland \newline
{\it Licensing provisions:} Standard CPC licence, \newline  {\tt http://cpc.cs.qub.ac.uk/licence/licence.html} \newline
{\it No. of lines in distributed program, including test data, etc.:} 10 888 \newline
{\it No. of bytes in distributed program, including test data, etc.:} 88 778 \newline
{\it Distribution format:}  tar.gz  \newline
{\it Programming language:} C. \newline
{\it Computer:} Laptop, Workstation. \newline
{\it Operating system:} Linux [tested on Linux (Kernel 2.6.9) with a gcc, version 3.4.6] \newline
{\it RAM:} Less than 10 MB  \newline
{\it Classification:} 17.15.  \newline
{\it Nature of problem:} The program calculates second-order energy moments, such as  variances and correlation coefficients, widely used as measures of the overall strength of an interaction and its similarity to other interactions. It allows for studies of the physical properties of various interactions and their effect on many-particle systems. \newline
{\it Solution method:} Calculations are based on spectral distribution theory  and invoke statistical measures provided by the theory. \newline
{\it Running time:} Less than 20  min (typically, several seconds)  using a 1.80 GHz processor.\newline
{\it References:} \\
J. B. French and K. F. Ratcliff, {\it Phys. Rev.} {\bf C 3}, 94 (1971); \\
F. S. Chang, J. B. French, and T. H. Thio, {\it Ann. Phys.} (N.Y.) {\bf 66}, 137 (1971); \\
K. T. Hecht and J. P. Draayer, {\it Nucl. Phys.} {\bf A223}, 285 (1974); \\
K. D. Sviratcheva, J. P. Draayer, and J. P. Vary, Nucl. Phys. A $\mathbf{786}$, 31 (2007).

\end{abstract}

\end{frontmatter}

\section{Introduction}
Spectral distribution theory (SDT) \cite{French66,FrenchR71,ChangFT71,HechtDraayer74} originated as a complement to conventional configuration-interaction spectroscopy (for a review, see \cite{KotaH10}). The efficacy of the theory stems from the finding that low-order energy moments of a microscopic interaction typically capture the dominant characteristic features of a many-particle system. While not the primary focus of SDT, convergence of the theory to exact eigensolution results improves, in principle,  as higher-order energy moments are taken into account, and/or in the limit of the many particles occupying a much larger available single-particle (s.p.) space. The theory also provides a relatively simple means to calculate important average properties, such as level densities, the degree of symmetry violation, and various other important features (e.g., \cite{French72,Draayer73,Parikh78,Kota79,CounteeDHK81,Potbhare77,French83, DraayerR83a, SarkarKK87,FrenchKPT88, KotaM94}).  In particular, in nuclear physics, the SDT approach  has been successfully applied to studies of energy spectra and reactions for $p$-, $sd$-, and $fp$-shell nuclei \cite{Ratcliff71,Ginnochio73,DraayerFPPW75, FrenchK83,StrohmaierGS87,AbzouziCZ91}, as well as for understanding dominant features and differences among $sd$-shell realistic effective interactions \cite{HalemaneKD78,DaltonBV79}. This has been achieved without the need for carrying out large-scale shell-model calculations, which are not always feasible. Recent applications include  explorations on quantum chaos, nuclear structure, and parity/time-reversal violation (for example, see \cite{BenetRW01,TomsovicJHB00,GomezKKMR03,HoroiGZ04,Kota05,Zhao05,TeranJ06,SviratchevaDV07,KotaH10}).

We focus on second-order energy moments, such as  variances and correlation coefficients, widely used as measures of the overall strength of an interaction and its similarity to other interactions. Furthermore, these moments can be propagated straightforwardly beyond the defining two-nucleon system to derivative systems with larger numbers of nucleons \cite{ChangFT71} and higher values of isospin \cite{HechtDraayer74}.  As a result, SDT allows one to gain further insight into the physical properties of various interactions and above all, into their effect on many-particle nuclear systems. For example, correlation coefficients can be used to extract important information on how well
pairing/rotational features develop in a nucleus given a specific interaction \cite{SDV06}. 

In addition, SDT gives an exact and simple prescription for identifying `density-dependent' monopole (centroid), one-body (induced single-particle energies), and its residual, irreducible two-body parts. The theory is readily extensible to 3-body interactions and beyond  and can be of special interest when such interactions are invoked, e.g., \cite{LisetskiyKBNSV09,HagenPDSNWP07,TsukiyamaBS11}. Hence, SDT framework  provides important information on the evolution of the shell structure, shell gaps, binding energies and contribution of 3-body forces with increasing number of particles \cite{HonmaOBM04,Otsuka01,LauneyDD12}.

In this paper, we present a detailed explanation of a computer code in C that utilizes SDT. The program is applicable to microscopic two-body nuclear interactions, regardless of whether they are realistic (like N$^3$LO  \cite{N3LO}, Argonne 18  \cite{AV18}, and CD-Bonn \cite{CDBonn}) or schematic (like pairing interaction), given in the $JT$-coupled basis of the  harmonic oscillator (HO) potential  or an arbitrary spherically symmetric potential. The program, as implemented, can be  straightforwardly modified to accommodate other coupling schemes (e.g., uncoupled, $m$-scheme, basis \cite{LauneyDD12}). 
The computer program is also extensible to 3-body interactions and beyond (see, e.g., Ref. \cite{LauneyDD12} for derivations in the $m$ scheme), and, in principle, can be expanded to include higher-order energy moments. 
The program introduces a new feature, namely, it handles the radial degeneracy of the `density-dependent'  one-body interactions and together with an efficient linked-list data structure, facilitates studies in large model spaces that go beyond valence-shell or two-shell applications of earlier Fortran programs \cite{Kota79,ChangDW82,GrimesBVH83}.  We also provide additional features that allow for studies of `density-dependent' interactions, as well as of interactions that can be expressed in terms of two other interactions that are mutually orthogonal. This, in turn, can provide, for example,  a key indication regarding how well a realistic nucleon-nucleon interaction may or may not reproduce -- without actually employing shell-model calculations -- prominent features of nuclei, such as pairing gaps in nuclear energy spectra or enhanced electric quadrupole transitions in collective rotational bands. The code is an essential computational tool that we have utilized in  recent explorations to provide important nuclear structure information \cite{SviratchevaDV07,SDV06,LauneyDD12} and can enable further studies of interest to nuclear theory (e.g., as suggested in \cite{CaprioLCHC12}).

\section{Spectral Distribution Theory -- notation and definitions}
Spectral distribution theory (or statistical spectroscopy) is well documented in the literature \cite{FrenchR71,ChangFT71,French72,HechtDraayer74,Kota79} and is accompanied by early computational codes \cite{Kota79,ChangDW82} for evaluating various measures. In this section, we specify the notation we use as well as present the formulae  used in the program to calculate the measures by following the appendix of Ref. \cite{SviratchevaDV07}. In addition, we include the radial degeneracy concept \cite{ChangFT71} to accommodate many HO major shells, together with a discussion on orthogonalizing an interaction with respect to a reference Hamiltonian and on expanding a Hamiltonian in terms of two other Hamiltonians.

In a standard second quantized form, a one- plus two-body isospin-conserving\footnote{Equivalently, the isospin-conserving part of a charge-dependent interaction (see Sec. \ref{programFeatures}).}  Hamiltonian is given
in terms of fermion creation $a_{\eta jm(1/2)\sigma }^\dagger$ and annihilation $\tilde a_{ \eta j-m(1/2)-\sigma } =  (-1)^{j-m+1/2 -\sigma }a_{ \eta jm(1/2)\sigma }$ tensors, where the operator $a^\dagger$ ($a$) creates (annihilates) a particle of type $\sigma =\pm 1/2$ (proton/neutron) in a state of oscillator quantum number $\eta$ ($\eta =0,\,1,\,2,\dots$ for $s$, $p$, $sd$, $\dots$ oscillator  shells, respectively), total angular momentum $j$ (half integer) with projection
$m$ in a finite space $2\Omega =\Sigma _j (2j+1)$,
\begin{eqnarray}
H&=&\sum _{r}\sqrt{[r]} \varepsilon _{r} 
\{a_r^\dagger \otimes \tilde a_r\}^{(00)} \nonumber\\
&&-\frac{1}{4}\sum_{rstu \Gamma}
\sqrt{(1+\delta _{rs})(1+\delta _{tu})[\Gamma]} W_{rstu}^\Gamma
\{\{a_r^\dagger \otimes a_s^\dagger\}^\Gamma \otimes
\{\tilde a_t \otimes \tilde a_u\}^\Gamma \}^{(00)},
\label{V2ndQF}
\end{eqnarray}
where the labels are $r=\{\eta_r,j_r ,\tau _r= \half\}$,
$[r]=2(2j_r+1)$, and $[\Gamma]=(2J+1)(2T+1)$. In Eq. (\ref{V2ndQF}),
$\varepsilon _{r}$ is an (external) single-particle energy and
$W_{rstu}^{JT} =\langle rsJTMT_0|H|tuJTMT_0 \rangle$ is a two-body antisymmetric 
matrix element  for $JT$-coupled normalized basis states, $|tuJTMT_0 \rangle = \frac{1}{\sqrt{1+\delta_{tu}} } \{a_t^\dagger \otimes a_u^\dagger\}^{JTMT_0} |0\rangle$,  
with
\begin{equation}
W_{rstu}^{\Gamma
}=-(-)^{r+s-\Gamma}W_{srtu}^{\Gamma }=
-(-)^{t+u-\Gamma } W_{rsut}^{\Gamma }=(-)^{r+s-t-u}W_{srut}^{\Gamma }=
W_{turs}^{\Gamma }.
\label{conjW}
\end{equation}

The underlying principle of SDT is the mapping of Hamiltonians onto a multi-dimensional linear vector space with an inner product. That is, a Hamiltonian $H$ can be realized as a vector with coordinates specified by the independent matrix elements of $H$.  For the one- plus two-body $H$, this space is spanned by a complete set of unit tensors (basis), $\{a_r^\dagger \otimes \tilde a_r\}^{(00)}$ and $\{\{a_r^\dagger \otimes a_s^\dagger\}^\Gamma \otimes \{\tilde a_t \otimes \tilde a_u\}^\Gamma \}^{(00)} $, and all such  Hamiltonians expand along  this basis according to Eq. (\ref{V2ndQF}). In this multi-dimensional space, an
inner product, $(H,H^\prime)$, 
for a pair of Hamiltonians $H$ and $H^\prime$,  is introduced (defined below), which, in turn, defines important measures; namely,  a norm of $H$, $\sigma_H=\sqrt{(H,H)}$ (``size" of a Hamiltonian), an angle between two Hamiltonians, $\cos \theta_{H,H^\prime}=(H,H^\prime)/\sigma_H \sigma_{H^\prime}$ (``orientation" of one Hamiltonian relative to the other), and a metric  (``distance"), 
$d(H,H^\prime)=\sigma_{(H-H^\prime)}$. For example, for the 3-dimensional coordinate space, an inner product is introduced by the scalar product of two vectors, $(\mathbf{x},\mathbf{x}^\prime)=\sum_{i=1}^3 x_i x^\prime_i={\rm Tr} (\mathbf{x} \mathbf{x}^\prime)$, which defines $\sigma_\mathbf{x}=\sqrt{x_1^2+x_2^2+x_3^2}$, the length of a vector, and a metric $d(\mathbf{x},\mathbf{x}^\prime)=\sqrt{\sum_{i=1}^3 (x_i -x^\prime_i)^2}$, the distance between two points in space. Note that these measures are independent of the basis representation, but are properties of the vectors.

The inner product in SDT is defined in terms of traceless operators,
$H -\langle H \rangle ^{(\alpha)}$,
where
\begin{equation}
\langle H \rangle ^{(\alpha)} = \frac{1}{\dimN_d} {\rm Tr}^{(\alpha)} H
\end{equation}
is the expectation value of $H$ averaged over the subset of basis states $\alpha$ (distributions) with dimensionality  $\dimN_d$ and thus, is related to the trace of the operator. E.g., for the matrix representation of a Hamiltonian with matrix elements $H_{fi}= \ME{f}{H}{i}$ for a final $\ket{f}$ and initial  $\ket{i}$ basis state, $\langle H \rangle ^{(\alpha)} =\sum_{i\subset {\alpha} } H_{ii}/\sum_{i\subset {\alpha} } 1$, and clearly, $H-\langle H \rangle^{(\alpha)}$ is traceless, such that 
${\rm Tr}^{(\alpha)}(H-\langle H \rangle^{(\alpha)})=\sum_{i\subset {\alpha} } H_{ii}-\langle H \rangle ^{(\alpha)} \sum_{i\subset {\alpha} } 1=0$.
We consider two distributions $\alpha $, namely, a {\it scalar} distribution (denoted by ``$n$" in
the formulae)  for a set of all states with a fixed $n$ particle number, as well as {\it isospin-scalar} distribution
(denoted by ``$n,T$") for a set of all states with fixed $n$ particle number  and  $T$ isospin 
value\footnote{A technical detail is that the model space in spectral distribution theory
is partitioned according to particular group symmetries and each
subsequent subgroup partitioning yields finer and more detailed
spectral estimates. Specifically,
for $n$ particles distributed over $4\Omega$ single-particle
states, the scalar
distribution averaged over all $n$-particle states is associated with
the \Un{}{4\Omega} group
structure 
and the isospin-scalar distribution averaged over the ensemble of all
$n$-particle states of isospin $T$ is associated with $\Un{}{2\Omega}
\otimes \Un{}{2}_T$.
}.
For a spectral distribution $\alpha $ ($\alpha $ is $n$ or $n,T$), the
inner product is defined as,
\begin{equation}
(H,H^\prime)^{(\alpha)}=\langle (H^\dagger -\langle H^\dagger \rangle ^\alpha ) (H^\prime-\langle H^\prime \rangle ^\alpha )
\rangle ^\alpha 
=\langle H^\dagger H^\prime\rangle ^\alpha -\langle H^\dagger \rangle
^\alpha \langle H^\prime \rangle ^\alpha.
\label{inner}
\end{equation}

Hence, the correlation coefficient between two Hamiltonian operators,
$H$ and $H^\prime$, is given as
\begin{equation}
\zeta ^{(\alpha)}
 _{H,H^\prime }= \frac{(H,H^\prime)^{(\alpha)}}{\sigma _H^{(\alpha)} \sigma _{H^\prime}^{(\alpha)}} 
=\frac{\langle H^\dagger H^\prime\rangle ^\alpha -\langle H^\dagger \rangle
^\alpha \langle H^\prime \rangle ^\alpha }{\sigma _H^{(\alpha)} \sigma _{H^\prime}^{(\alpha)}},
\label{zeta}
\end{equation}
where the ``width" of $H$ (related to the Hilbert-Schmidt norm) is the positive square root of the
variance,
\begin{equation}
(\sigma ^{(\alpha)}
 _{H})^2=(H,H)^{(\alpha)}
=\langle H^2 \rangle ^\alpha -(\langle H \rangle ^\alpha )^2.
\label{sigma}
\end{equation}
The steps for computing these quantities are outlined in Sections \ref{secSPE} and \ref{secTBME}. The significance
of a positive correlation coefficient is given by Cohen \cite{Cohen88_03} and
later revised to the following, $\zeta = 0.00-0.09$ represents a trivial correlation, 
$\zeta = 0.10-0.29$, $0.30-0.49$, $0.50-0.69$, and  $ 0.70-0.89$  represent small,  medium,  large and very large correlations, respectively,
while $\zeta =0.90-0.99 $ and $1.00$ represent  nearly perfect and
perfect correlations, respectively.

\begin{figure}[th]
\begin{minipage}{0.52\textwidth}
\begin{center}
\includegraphics[width=0.65\textwidth]{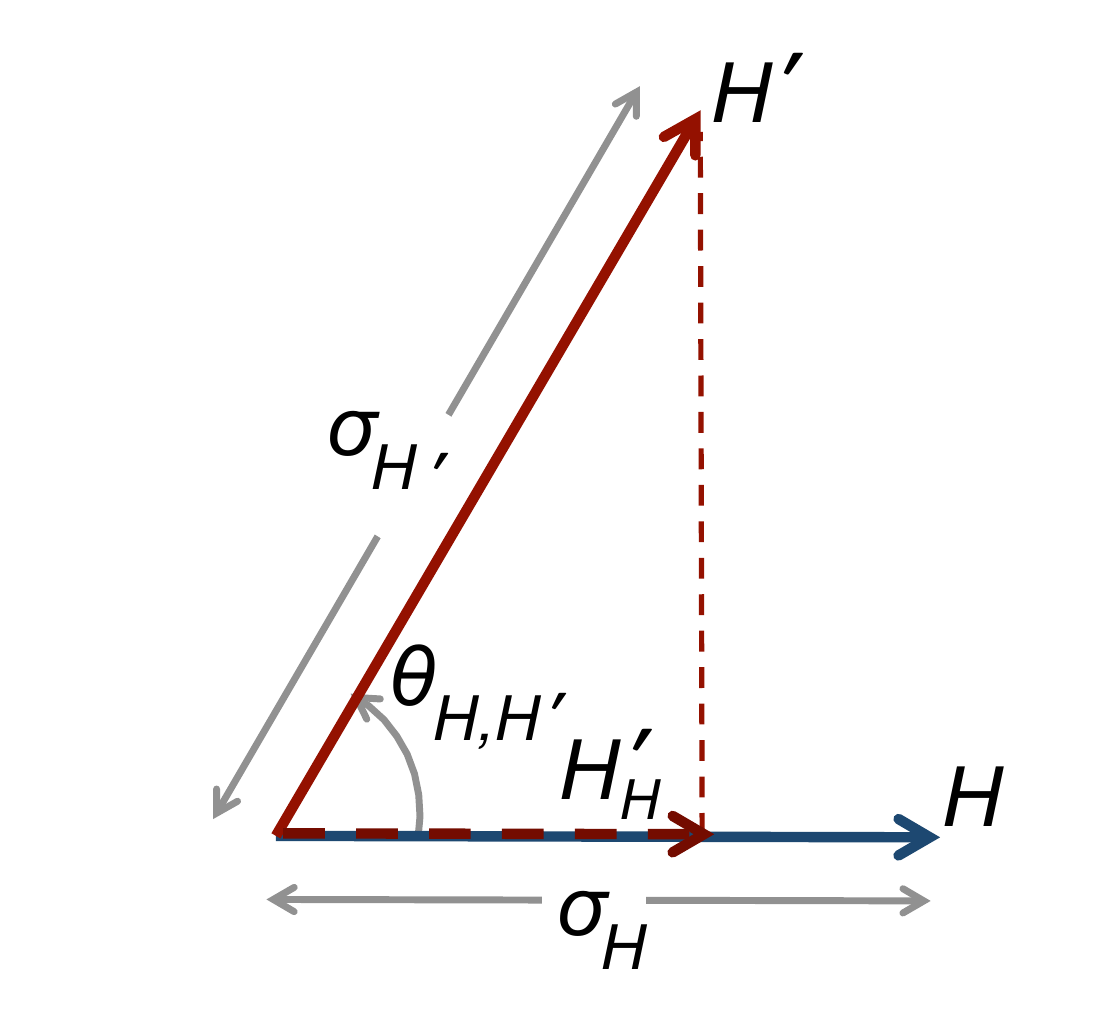}
\end{center}
\end{minipage}
\begin{minipage}{0.48\textwidth}
\caption{
Geometrical interpretation of two Hamiltonians in the framework of spectral distribution theory. Each  Hamiltonian ($H$ and $H^\prime$) is represented by a vector of length given by its respective norm $\sigma $; the angle between $H$ and $H^\prime$ is specified by the corresponding correlation coefficient, $\theta_{H,H^\prime} = \arccos \zeta_{H,H^\prime}$. $H^\prime_H$ is the projection of $H^\prime$ onto $H$ and reflects the spectral characteristics of $H$.
 \label{sdt_viz_vecs}
 }
\end{minipage}
\end{figure}

From a geometrical perspective, in spectral distribution theory every interaction is associated with a vector (Fig. \ref{sdt_viz_vecs}) of length  $\sigma$ (Eq. \ref{sigma}) and the correlation coefficient $\zeta $ (Eq. \ref{zeta}) defines the angle (via a normalized scalar product) between two vectors. Hence, $\zeta _{H,H^\prime}$
gives the normalized projection of $H^\prime$ onto the $H$ interaction (or $H$ onto $H^\prime$). In addition, $(\zeta _{H,H^\prime})^2$ gives the percentage of $H^\prime$ that reflects the characteristic properties of the $H$ interaction. As for $\sigma_H$, it is a natural measure of the $H$ operator size and  realizes the spread of the $H$ eigenvalue distribution. As is well-known, the smaller the $\sigma_H$ (the weaker the interaction), the more compressed the energy spectrum for $H$ \cite{FrenchR71}. Ê

As mentioned above, $H$  and $H^\prime$  of Fig. \ref{sdt_viz_vecs} lie in a multi-dimensional space, spanned by a complete set of unit tensors.
This allows one to identify, within this multi-dimensional space, a set of Hamiltonians $\{H\}$ (multi-dimensional vector $H$ in Fig. \ref{sdt_viz_vecs}) and then project any $H^\prime$ onto the space spanned by $\{H\}$, thereby providing an expansion of $H^\prime$ in terms of other Hamiltonians,  $H^\prime=\oplus H + H_{\rm res}$. Here, $H_{\rm res}$ is the part of $H^\prime$ that lies outside the $\{H\}$ space (multi-dimensional dashed line in Fig. \ref{sdt_viz_vecs}). In this study, we focus on the case of a two-dimensional $\{H\}$ space (Section \ref{orthogH}). 

The similarity between two interactions can be further tracked in many-nucleon systems
\cite{HechtDraayer74,KotaPP80} through propagation
formulae. The latter
determine how the  averages
extracted from the two-nucleon matrix elements get carried forward
into many-nucleon systems. This
propagation of information is model-independent.
In order to calculate energy moments and their propagation in many-particle systems, that is, for higher $n$
(and $T$) values, each Hamiltonian $H$, which consists of one- ($k=1$) and
two-body ($k=2$) interactions, is expressed as a linear combination
of terms of
definite particle rank (irreducible tensors ${\mathcal H}_k(\nu )$ of rank $\nu
=0,1,2$), that is as a collection of pure zero-, one- and two-body
interactions. For example, in the case of scalar distribution and for
$n$ particles \cite{French72}, the Hamiltonian can be 
rendered\footnote{
As explained by French \cite{French72}, for a scalar distribution ($\alpha=n$), a 2-body Hamiltonian can be constructed in terms of simpler 0- and 1-body terms with the use of the one-body operator $n$. E.g., with $(n-1)$ being (0+1)-body, if one finds a pure 1-body ${\mathcal H}_2(1)$ [that vanishes in the  zero-particle space, ${\mathcal H}_2(1)|0\rangle=0$], $(n-1){\mathcal H}_2(1)$ can only be, in general, (1+2)-body -- and as it  vanishes in the one-particle space, for $n=1$, it is in fact 2-body;  similarly for $\binom{n}{2} {\mathcal H}_2(0)$ (it vanishes for $n=0$ and $n=1$). Hence, the generalization for constructing a $k$-body interaction in terms of pure particle-rank interactions is given as,
$H=\sum_{\nu=0}^k \binom{n-\nu}{k-\nu}{\mathcal H}_k(\nu)=\sum_{\nu=0}^k \frac{(n-\nu)(n-\nu-1)\dots (n-k+1)}{(k-\nu)!}{\mathcal H}_k(\nu)$ -- clearly, it vanishes for $n=\nu,\nu+1,\dots,k-1$ due to the $n$-dependent polynomial and for $n<\nu$ due to the $\nu$-body ${\mathcal H}_k(\nu)$. Note that for an isospin-scalar distribution ($\alpha=n,T$), two operators are invoked, $n$ and ${\mathbf T}^2$, such that, e.g., $ \frac{n(n+2)-4 {\mathbf T}^2}{8} {\mathcal H}_{2}^{T=0}(0)$ together with $ \frac{3 n(n-2)+4 {\mathbf T}^2}{8}{\mathcal H}_{2}^{T=1}(0)$ contribute to a 2-body Hamiltonian (they vanish for $n=0, T=0$ and $n=1,T=1/2$).
},
\begin{equation}
H=n{\mathcal H}_1(0)+\binom{n}{2} {\mathcal H}_2(0)+
{\mathcal H}_1(1)+(n-1){\mathcal H}_2(1) +{\mathcal H}_2(2).
\label{HtoPureH}
\end{equation}
For then the inner product (Eq. \ref{inner}) that defines the correlation coefficient
(Eq. \ref{zeta}) or the variance  (Eq. \ref{sigma}) is easily
computed for different particle numbers $n$ using the irreducible tensors and their expectation values in a many-particle basis,
\begin{eqnarray}
\zeta ^{(\alpha)}
 _{H,H^\prime }\sigma^{(\alpha)}
 _H \sigma^{(\alpha)}
 _{H^\prime} 
&=& \langle [{\mathcal H^\dagger}_1(1)+(n-1){\mathcal H^\dagger}_2(1)] [{\mathcal H'}_1(1)+(n-1){\mathcal H'}_2(1) ]\rangle ^\alpha  \nonumber\\
&+&\langle {\mathcal H^\dagger}_2(2) {\mathcal H'}_2(2) \rangle ^\alpha, \\
(\sigma^{(\alpha)}
 _H )^2
&=& \langle [{\mathcal H}_1(1)+(n-1){\mathcal H}_2(1)]^2 \rangle ^\alpha  + \langle {[\mathcal H}_2(2)]^2 \rangle ^\alpha. 
\label{zeta_pure}
\end{eqnarray}

\subsection{One-body interaction  and single-particle energies SPE's \label{secSPE}}
For a set $s$ of s.p. states (typically, referred to as orbits) of degeneracy $\dimN_s$, the average single-particle energy $\varepsilon$ and
the traceless single-particle energy, $\tilde \varepsilon_r$, of the $r^{\text{th}}$ orbit are given as,
\sublabon{equation}
\begin{eqnarray}
&&\varepsilon =\frac{1}{\dimN}\sum _s \varepsilon _s \dimN_s \label{espe0}\\
&&\tilde \varepsilon _r= \varepsilon_r - \varepsilon = \varepsilon _r- \frac{1}{\dimN}\sum _s \varepsilon _s \dimN_s, \label{espe1}
\end{eqnarray}
\sublaboff{equation}
where $\dimN=\sum _s \dimN_s$ and for a $jj$-coupled HO s.p. basis, $\dimN_s=2\sum _s(2j_s+1)$. A quick check for $\tilde \varepsilon$, ${\rm Tr} \tilde \varepsilon=\sum _r \dimN_r \tilde \varepsilon_r =\sum _r \dimN_r (\varepsilon_r - \varepsilon)=\dimN \varepsilon - \varepsilon \sum _r \dimN_r=0$, confirms that it is traceless.

These quantities construct pure 0-body and 1-body terms (irreducible interaction tensors of particle ranks 0 and 1) for a one-body interaction, namely, 
\begin{eqnarray}
{\mathcal H}_1(0) &=& \varepsilon \nonumber\\
{\mathcal H}_1(1) &=& \sum _{r}\sqrt{[r]} \tilde \varepsilon _{r}\{a_r^\dagger \otimes \tilde a_r\}^{(00)}. 
\nonumber
\end{eqnarray}
With these definitions and using that $n=\sum _{r}\sqrt{[r]}\{a_r^\dagger \otimes \tilde a_r\}^{(00)}$, one can verify that a one-body $H=n{\mathcal H}_1(0)+{\mathcal H}_1(1)$, as shown in (\ref{HtoPureH}).

\subsection{Two-body interaction and two-body matrix elements TBME's \label{secTBME}}
\noindent
{\it Scalar distribution. -- } In this case, degeneracy dimensionalities are given as,
\begin{equation}
\dimN=\sum _r \dimN_r\,{\rm , where  }\,\dimN_r=2(2j_r+1).
\label{dimSc}
\end{equation}
For a two-particle system, the monopole
moment (centroid), which
is the average expectation value of the two-body interaction,  is
defined as
\begin{equation}
V_c=\frac{1}{\binom{\dimN}{2}}\sum _{r \le s,
\Gamma}[\Gamma] W_{rsrs}^\Gamma,
\label{WcS}
\end{equation}
where the $\Gamma$-sum
goes over all
possible $(J,T)$ for given $r$ and $s$, while $\binom{\dimN}{2}=\sum _{r \le s,
\Gamma}[\Gamma]$.
The traceless induced single-particle energy  is constructed as 
(that is, by contraction of the two-body interaction into an effective one-body operator under the
particular group structure),
\begin{equation}
\lambda _{rt} =\frac{1}{\dimN_r}\sum _{s, \Gamma} [\Gamma]
W_{rsts}^{\Gamma} \sqrt{(1+\delta _{rs}) (1+\delta _{ts})}\hat \delta_{rt}
-\frac{\delta_{rt}}{\dimN} \sum _{tu, \Gamma} [\Gamma]
W_{tutu}^{\Gamma} (1+\delta _{tu}),
\label{ispeS}
\end{equation}
where $\hat \delta_{rt}$ is 1 if $j_r=j_t$ and $(-1)^{\eta_r}=(-1)^{\eta_t}$. We thus take into account
a radial degeneracy \cite{ChangFT71,FrenchR71}, that is, not every s.p. state is distinguishable by the $j$ angular momentum and parity, thereby leading to an effective 1-body interaction that mixes different oscillator quantum numbers (equivalently, radial quantum numbers). This, for example, introduces a nonzero $\lambda _{rt}$ with $r=\{\eta_r=0,j_r=\half ,\tau _r= \half\}$ and $t=\{\eta_t=2,j_t=\half ,\tau _t= \half\}$.

In turn, the traceless pure two-body interaction is defined through the antisymmetrized matrix elements,
\begin{equation}
W_{rstu}^{\Gamma}(2)=W_{rstu}^{\Gamma}-V_c \delta _{rt}\delta _{su}
-\frac{
\lambda _{rt}\delta_{su} +\lambda _{su}\delta_{rt}-(-1)^{r+s-\Gamma}(\lambda _{st}\delta_{ru} + \lambda _{ru}\delta_{st})}
{(\dimN-2) \sqrt{(1+\delta _{rs}) (1+\delta _{ts})} }.
\label{W2S}
\end{equation}

The quantities defined above, $V_c$, $\lambda$, and $W(2)$, specify the corresponding
 tensors of particle ranks 0, 1, and 2 for a two-body interaction, namely, 
\begin{eqnarray}
{\mathcal H}_2(0) &=& V_c,  \nonumber\\
{\mathcal H}_2(1)&=& \sum _{rs}\sqrt{[r]} \frac{\lambda _{rs}}{\dimN-2}\{a_r^\dagger
\otimes \tilde a_s\}^{(00)},  \\
{\mathcal H}_2(2) &=& 
\textstyle{\sum ^*}
\frac{\sqrt{[\Gamma]}}{\sqrt{(1+\delta _{rs})(1+\delta _{tu})}}
W_{rstu}^\Gamma (2)\{\{a_r^\dagger \otimes a_s^\dagger\}^\Gamma \otimes
\{\tilde a_t \otimes \tilde a_u\}^\Gamma \}^{(00)},
\nonumber
\end{eqnarray}
where the sum $\sum^*$ goes over $r\leq s$, $t\leq u$ and $\Gamma=(J,T)$.

Hence, the quantity that defines the correlation coefficient (Eq. \ref{zeta}) and the norm  for $H=H^\prime$ (Eq. \ref{sigma}) is easily
computed for a larger number of particles, $n$, 
\sublabon{equation}
\begin{eqnarray}
&& \zeta ^{(n)} _{H,H^\prime }\sigma^{(n)}_H \sigma^{(n)}_{H^\prime} = \langle H^\dagger H^\prime\rangle ^n-\langle H^\dagger \rangle ^n\langle H^\prime \rangle ^n  = \nonumber \\ 
&& \frac{n(\dimN-n)}{(\dimN-1)} \left( \frac{1}{\dimN}\sum _{r} \dimN_r \tilde \varepsilon _{r} \tilde \varepsilon ^\prime _{r} \right) + \label{<HH'>nspe} \\ 
&& \frac{n(\dimN-n)(n-1)}{(\dimN-1)(\dimN-2)} 
\left(\frac{1}{\dimN} \sum _{r} \dimN_r (\tilde \varepsilon _{r} \lambda ^\prime _{r}  +\lambda _{r} \tilde \varepsilon ^\prime _{r} ) \right) + \label{<HH'>nspelm} \\ 
&& \frac{n(\dimN-n)(n-1)^2}{(\dimN-1)(\dimN-2)^2} 
\left(\frac{1}{\dimN}  \sum _{rs} \dimN_r \lambda _{rs} \lambda ^\prime _{rs} \right) + \label{<HH'>nlm} \\ 
&&\frac{n(n-1)(\dimN-n)(\dimN-n-1)}{2(\dimN-2)(\dimN-3)} 
\left(\frac{1}{\binom{\dimN}{2}}
\textstyle{\sum ^*}
[\Gamma] W_{rstu}^\Gamma (2) {W^\prime }_{rstu}^\Gamma (2) \right). \label{<HH'>n2b} 
\end{eqnarray}
\sublaboff{equation}

\noindent
{\it Isospin-scalar distribution. -- } In this case, degeneracy dimensionalities are given as,
\begin{equation}
\dimN=\sum _r \dimN_r\,{\rm , where  }\,\dimN_r=2j_r+1,
\label{dimSc}
\end{equation}
and the centroid is,
\begin{equation}
V_c^T=\frac{2}{\dimN (\dimN+(-1)^T)}\sum _{r \le s, J}[J]
W_{rsrs}^{JT}.
\label{WcTS}
\end{equation}
The $\lambda _{rt} ^T$ traceless induced
single-particle energies, which specify ${\mathcal H}_2(1)$, and the $W_{rstu}^{JT}(2)$ two-body matrix elements  \cite{Kota79} that specify the traceless pure 2-body interaction ${\mathcal H}_2(2)$
are defined as,
\begin{equation}
\lambda _{rt} ^T = \frac{1}{\dimN_r}\sum _{s,J} [J] W_{rsts}^{JT} \sqrt{(1+\delta _{rs}) (1+\delta _{ts})}\hat \delta_{rt}
-\frac{\delta_{rt}}{\dimN} \sum _{tu,J} [J]
W_{tutu}^{JT} (1+\delta _{tu}), \label{ispeTS} 
\end{equation}
\begin{eqnarray}
W_{rstu}^{JT}(2) &=& W_{rstu}^{JT}-V_c^T \delta _{rt}\delta _{su} - \nonumber \\
&&\frac{\lambda^T _{rt}\delta_{su} +\lambda^T _{su}\delta_{rt}-(-1)^{r+s-J-T}(\lambda^T _{st}\delta_{ru} + \lambda^T _{ru}\delta_{st})}
{({\dimN+2(-1)^T}) \sqrt{(1+\delta _{rs}) (1+\delta _{ts})} }. \label{W2TS}
\end{eqnarray}
The quantities defined above are used to calculate the correlation coefficient $\zeta ^{(n,T)}$ (Eq. \ref{zeta})  and the
variance $(\sigma ^{(n,T)})^2$ for $H=H^\prime$ (Eq. \ref{sigma}) for higher values of $n$ and $T$, by employing
\sublabon{equation}
\begin{eqnarray}
&& \zeta ^{(n,T)} _{H,H^\prime }\sigma^{(n,T)}_H \sigma^{(n,T)}_{H^\prime} =\langle H^\dagger H^\prime\rangle ^{n,T}-\langle H^\dagger \rangle ^{n,T}\langle H^\prime \rangle ^{n,T}= \nonumber \\
&& {\mathcal P}_1(n,T) \left( \frac{1}{\dimN}\sum _{r} \dimN_r \tilde \varepsilon _{r} \tilde \varepsilon ^\prime _{r} \right) + \label{<HH'>nTspe} \\ 
&&\sum _{\tau } {\mathcal P}_1(n,T,\tau )
\left(\frac{1}{\dimN} \sum _{r} \dimN_r (\tilde \varepsilon _{r} \lambda ^\prime _{r}  +\lambda _{r} \tilde \varepsilon ^\prime _{r} ) \right) + \label{<HH'>nTspelm} \\ 
&&\sum _{\{\tau _1,\tau _2 \}} {\mathcal P}_1(n,T,\tau _1,\tau _2)
\left(\frac{1}{\dimN}  \sum _{rs} \dimN_r \half [\lambda _{rs}^{\tau _1}{\lambda ^\prime }_{rs}^{\tau _2} +
      \lambda _{rs}^{\tau _2} {\lambda ^\prime }_{rs} ^{\tau _1}] \label{<HH'>nT} \right) + \label{<HH'>nTlm} \\ 
&&\sum _\tau  {\mathcal P}_2(n,T,\tau ) \left( \frac{2}{\dimN (\dimN+(-1)^\tau )}
\textstyle{\sum ^*}
[J] W_{rstu}^{J\tau } (2) {W^\prime }_{rstu}^{J\tau } (2) \right), \label{<HH'>nT2b} 
\end{eqnarray}
\sublaboff{equation}

\noindent
where $\tau $ is $0$ or $1$, and the set $\{\tau _1,\tau _2\}$
is $\{0,0\}, \{0,1\}$ or $\{1,1\}$. The  sum $\sum^*$ goes over $r\leq s$,
$t\leq u$ and $J$. The propagator functions are derived in
\cite{French69,HechtDraayer74} and shown below for completeness:
\begin{align}
&{\mathcal P}_1(n,T)=\textstyle{
\frac{n(\dimN +2)(\dimN-\frac{n}{2})-2 \dimN T(T+1)}
{(\dimN-1)(\dimN +1)}
},
\label{propagators} \\
&{\mathcal P}_1(n,T,\tau )=
\textstyle{
\frac{
4\dimN T(T+1) (1-n) (1-(-1)^\tau )+
(\dimN +2)({\mathcal N}-\frac{n}{2}) [(2\tau 
+1)n(n+2(-1)^\tau)-4T(T+1)(-1)^\tau ]
}{4 (\dimN -1)(\dimN +1)(\dimN +2(-1)^\tau )}
},
\nonumber
\\
&{\mathcal P}_1(n,T,\tau _1={\rm odd},\tau _2)= \nonumber \\
&\textstyle{
\frac{8\dimN T(T+1) (n-1) (\dimN - 2n +4)+[(2\tau _1+1)n(n-2)+4T(T+1)] [\frac{(2{\tau _2}+1)(n+2(-1)^{\tau _2})}{2}(\dimN -\frac{n}{2})+T(T+1)(-1)^{\tau _2}]
[\dimN-2(-1)^{\tau _2}]}{8(\dimN -1)(\dimN +1)(\dimN -2)^2}
},
\nonumber
\\
&{\mathcal P}_1(n,T,\tau _1={\rm even},\tau _2)= \nonumber \\
&\textstyle{
\frac{[(2\tau _1+1)n(n+2)-4T(T+1)] [\frac{(2{\tau _2}+1)(n+2(-1)^{\tau _2})}{2}(\dimN -\frac{n}{2})+T(T+1)(-1)^{\tau _2}]
[\dimN-2(-1)^{\tau _2}]}{8(\dimN -1)(\dimN +1)(\dimN -2)(\dimN +2)}
},
\nonumber
\\
&{\mathcal P}_2(n,T,\tau =0)= \textstyle{\frac{[n(n+2)-4T(T+1)]
[(\dimN -\frac{n}{2})(\dimN -\frac{n}{2}+1)-T(T+1)]}{8\dimN (\dimN-1)}
},
\nonumber
\\
&{\mathcal P}_2(n,T,\tau =1)= 
\textstyle{
\frac{
 T^2 (T+1)^2\frac{3\dimN ^2-7\dimN +6}{2}
+\frac{3n(n-2)}{8}(\dimN -\frac{n}{2})(\dimN -\frac{n}{2}-1)(\dimN
+1)(\dimN +2)
}
{\dimN (\dimN +1)(\dimN -2)(\dimN -3)}+
}\nonumber \\
&\hspace{1.18in}\textstyle{
\frac{
\frac{T(T+1)}{2}[(5\dimN -3)(\dimN +2)n(\frac{n}{2}-\dimN) +
\dimN (\dimN -1)(\dimN +1)(\dimN +6)]
}
{\dimN (\dimN +1)(\dimN -2)(\dimN -3)}
}. \nonumber
\end{align}

\subsection{Orthogonal Hamiltonians \label{orthogH}}
Orthogonal Hamiltonians are important in SDT, as they provide an orthogonal ``basis", along which a  Hamiltonian in consideration, $K$, can be projected. 
Large overlaps of $K$ with a ``basis" Hamiltonian implies that features of the latter are also well realized in $K$, while the orthogonality ensures that other ``basis" Hamiltonians point to complementary features. For example, if $K$ projects onto a $Q\cdot Q$ interaction and an $L^2$ interaction on an equal footing, this  does not necessarily imply that both $Q\cdot Q$  and $L^2$ are required to describe related features of $K$, as $Q\cdot Q$  and $L^2$ can indeed highly overlap. 

\begin{figure}[th]
\begin{minipage}{0.6\textwidth}
\includegraphics[width=\textwidth]{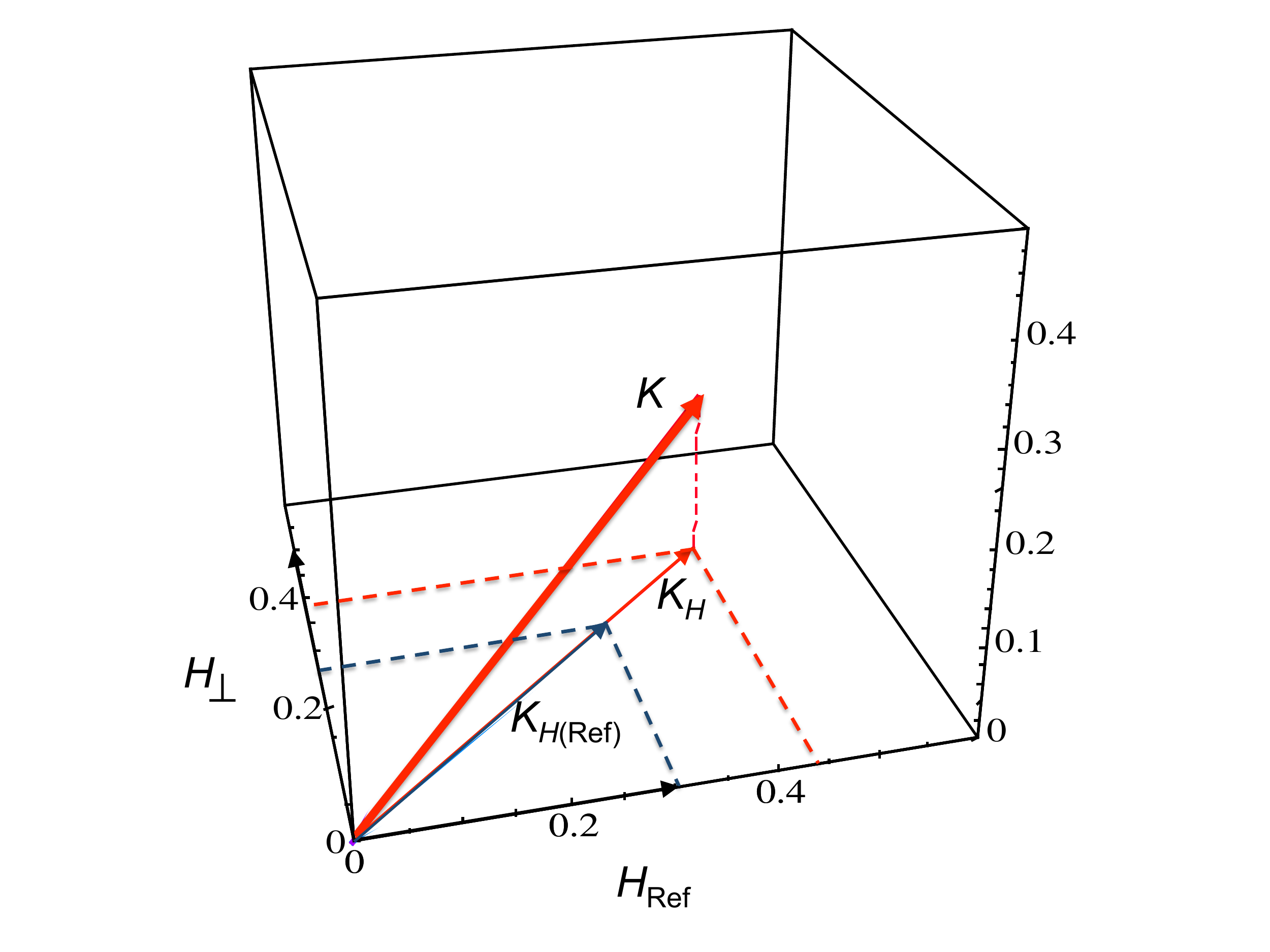}
\end{minipage}
\begin{minipage}{0.4\textwidth}
\caption{
Geometrical interpretation of expanding a Hamiltonian $K$ in terms of two Hamiltonians $H_{\rm Ref} $ and $H_{\perp}$, that is,  projecting  a vector $K$ onto a plane defined by the two `orthogonal vectors'. $K_H$ is the projection of $K$ onto the plane, while $K_{H {\rm (Ref)}}$ is the portion that does not involve $H_{\rm Ref}$ renormalization.
\label{sdt_viz}}
\end{minipage}
\end{figure}

In the SDT framework, as shown in Ref. \cite{Potbhare77}, one can straightforwardly extract  the traceless 2-body $H_{\perp}$ orthogonal part of a Hamiltonian $H$ with respect to a reference one, $H_{\rm Ref}$, such that the condition $\langle H_{\perp}^\dagger H_{\rm Ref}\rangle=0$ is satisfied. This is done  in a way analogous to a Gram-Schmidt orthogonalization of two vectors $\vec h$ (reference) and $\vec g$, so that a new vector orthogonal to $\vec h$ is given as, 
$\vec g_{\perp}=\vec g-\frac{\vec g \cdot \vec h}{\vec h \cdot \vec h}\vec h=\vec g-\frac{\zeta_{g,h}\sigma_{g}}{\sigma_h}\vec h$. Similarly, the traceless 2-body $H_{\perp}$ is constructed using, 
\begin{equation}
{W^{JT}_{rstu}(2)}_H-\frac{\zeta_{H,H_{\rm Ref}} \sigma_{H} }{ \sigma_{H_{\rm Ref}} }{W^{JT}_{rstu}(2)}_{H_{\rm Ref }}.
\label{orthogTr0}
\end{equation}
Then, as illustrated in Fig. \ref{sdt_viz}, any interaction $K$ can be projected ($K_H$) \cite{DraayerR82} onto a plane defined by the two `orthogonal vectors', $K_H=c_{ H_{\rm Ref} } H_{\rm Ref} + c_{ H_{\perp} } H_{\perp}$ ($K=K_H+K_{\rm res}$ with a residual interaction $K_{\rm res}$ perpendicular to this plane), where the projection coefficients are given by,
\sublabon{equation}
\begin{eqnarray}
c_{H_{\perp}} &=& \zeta_{K,H_{\perp}}\sigma_K/\sigma_{H_{\perp}}  \label{xLofHR} \\
c_{H_{\rm Ref}} &=& \zeta_{K,H_{\rm Ref}}\sigma_K/\sigma_{H_{\rm Ref}},  \label{xREFofHR}
\end{eqnarray}
and the cosine of the angle between $K$ and the plane (or equally, and $K_H$) is,
\begin{equation}
\zeta_{K,K_H}=\sqrt{  \zeta^2_{K,H_{\rm Ref}}+\zeta^2_{K,H_{\perp}} }.
\label{cc_proj}
\end{equation}
The norm of $K_H$ can be then calculated as $\sigma_{ K_H}=\sigma_{ K}\zeta_{K,K_H}$. A more interesting case is a normalized projection of $K$ (which preserves the norm of $H_{\rm Ref}$, see Fig. \ref{sdt_viz}), $K_{H {\rm (Ref)}}=\frac{K_H}{c_{ H_{\rm Ref} } }= H_{\rm Ref} + \frac{c_{ H_{\perp} }}{  c_{ H_{\rm Ref} }  } H_{\perp}$, with a norm,
\begin{equation}
\sigma_{K_{H\rm (Ref)}} = \sigma_{H_{\rm Ref}}  \sqrt{ \left(1+ \zeta^2_{K,H_{\perp}} / \zeta^2_{K,H_{\rm Ref}} \right) }.
\label{norm_proj}
\end{equation}
\sublaboff{equation}
This allows for studies that utilize $H_{\rm Ref}$ that has been already deduced from selected nuclear properties, such as a standard pairing Hamiltonian with a pairing strength adjusted to reproduce pairing gaps in nuclei (also see \cite{SDV06}). 

\section{Program Performance \label{programFeatures}}
According to the current default (see {\tt  $<$calculateMomentsDr.h$>$}), the program can handle up to 45 single-$j$ orbits (that is, 9 major HO shells, $\eta=0,\, 1,\dots,8$), as well as up to 20 Hamiltonians that are simultaneously considered in a given calculation. These limits can be increased if an application requires more orbits and/or a larger set of Hamiltonians. In addition, a subset of orbits can be selected.
\begin{figure}[th]
\begin{center}
\includegraphics[width=0.42\textwidth]{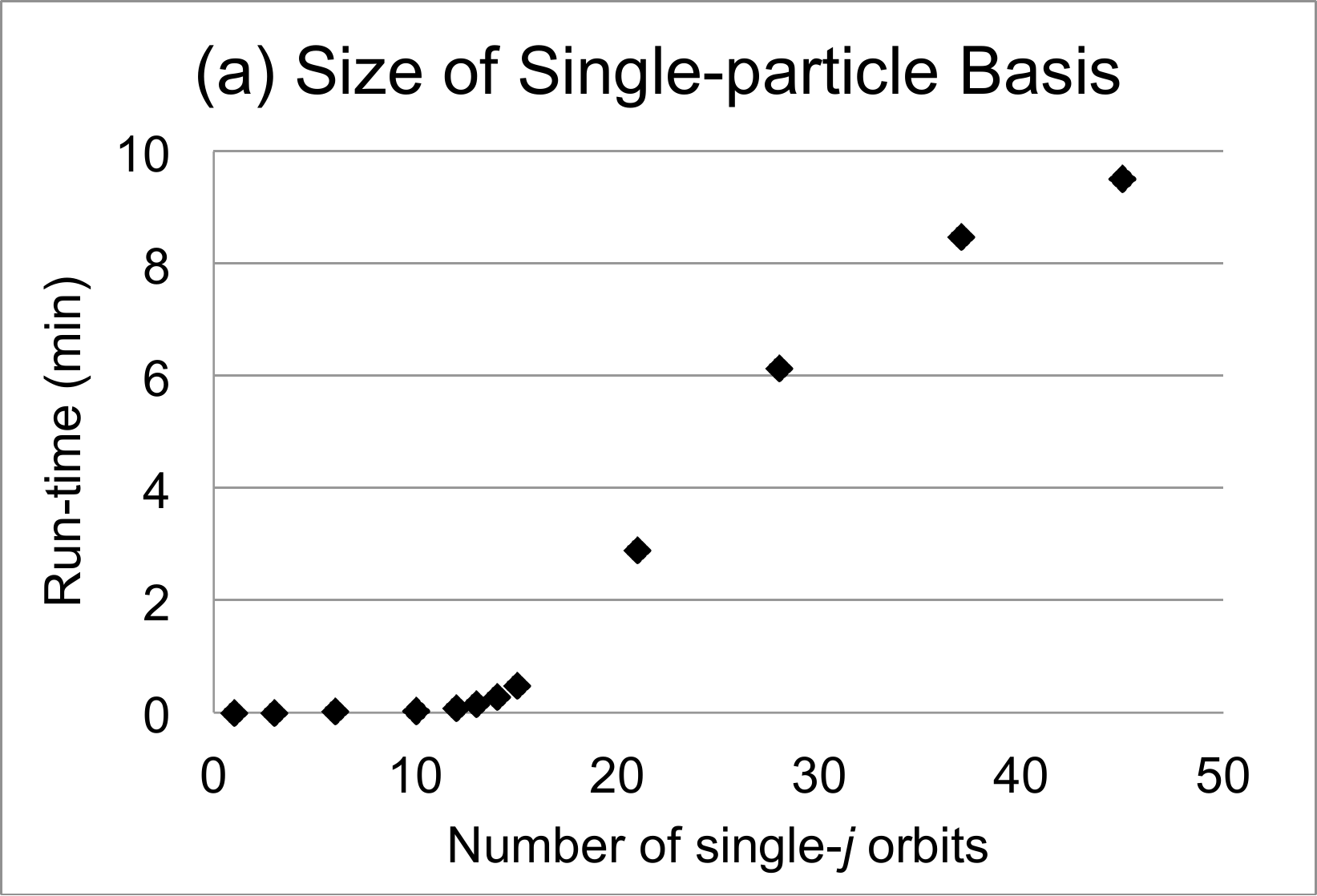}
\includegraphics[width=0.42\textwidth]{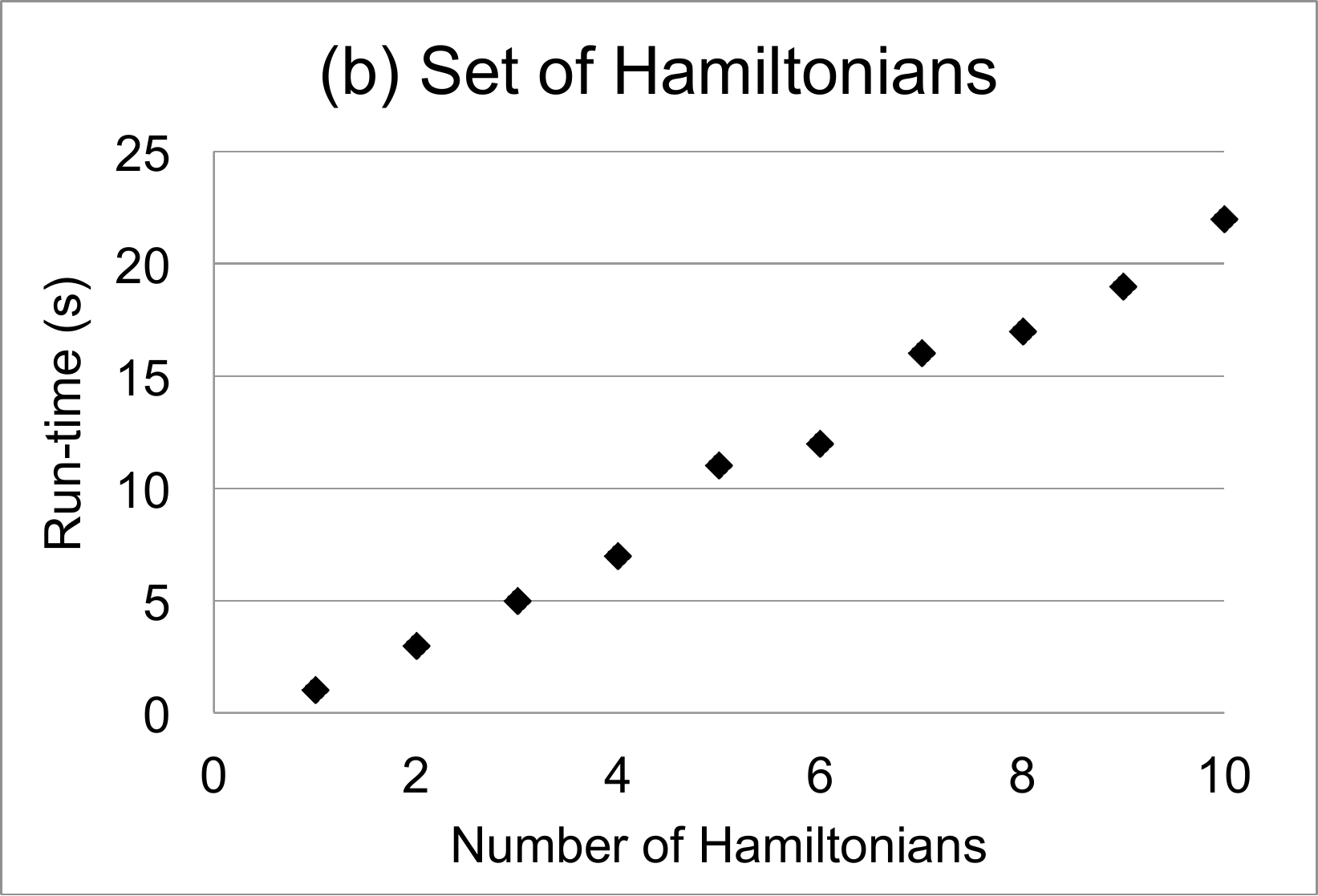}
\caption{Run-time as a function of (a) the size of the s.p. basis using 3 Hamiltonians  and (b) the number of Hamiltonians that are simultaneously considered in a given calculation employing 15 single-$j$ orbits.}
\label{test_runtime}
\end{center}
\end{figure}

The computer program has been applied to various interactions including the realistic interactions N$^3$LO \cite{N3LO}, JISP16 \cite{JISP16}, CD-Bonn \cite{CDBonn} and AV18 \cite{AV18}, as well as their effective counterparts. Only the isospin-conserving part,
\begin{eqnarray}
W_{rstu}^{JT(IC)}=\left\{
\begin{array}{ll}
W_{rstu}^{JT(pn)}  &,\, T=0    \\
\frac{1}{3}(W_{rstu}^{JT(pp)}+W_{rstu}^{JT(nn)}+W_{rstu}^{JT(pn)})  &,\, T=1     \\
\end{array}
\right.
\end{eqnarray}
has been considered for charge-dependent interactions, that is, with different  $pp$ $(T_0=1)$, $nn$ $(T_0=-1)$, and $pn$ $(T_0=0)$ $T=1$ matrix elements, $W_{rstu}^{JT(T_0)}=\langle rsJTMT_0|H|tuJTMT_0 \rangle$. The much smaller isospin-mixing part of nuclear interactions cannot be treated in the current program, but could be included, as the associated formulae have been already derived  in Ref. \cite{HechtDraayer74}. In addition, several other interactions have been utilized, e.g., GXPF1 \cite{HonmaOBM04}, as well as conventional $pn$- plus like-particle pairing and $Q\cdot Q$ interactions, and the Sp(4) pairing Hamiltonian \cite{SviratchevaGD04}. Interactions may be specified for a large set of orbits and only a subset of orbits can be selected in the program.

The program run-time is studied as a function of the number of s.p. states, as well as the number of Hamiltonians considered (Fig. \ref{test_runtime}). While the run-time increases exponentially, as expected, for larger model spaces (more orbits considered), it remains reasonably small even when major shells up to the $sdgik$ shell are considered (45 single-$j$ orbits). For a broad range of problems of interest that explore up through the $sdg$ shell (15 orbits), the program is indeed very fast (Fig. \ref{test_runtime}a). Furthermore, the run-time is only linearly dependent on the number of Hamiltonians considered, as shown in Fig. \ref{test_runtime}b for 15 orbits. This makes the program easily applicable to a large set of Hamiltonians, which in turn benefits calculations of projecting a given interaction $K$ simultaneously along many distinct interactions $\{H\}$, especially when symmetry-preserving terms are involved. Such a project is currently being implemented as an extension of the present computational code.

Finally, the program benefits from an efficient way for storing TBME's, namely, matrix elements are stored in linked lists that are grouped, while being read, according to the role each $W^{JT}_{rstu}$ plays  in various sums that are needed to construct irreducible tensors. For example, one group consists of all $W^{JT}_{rsrs}$ and another group consists of all $W^{JT}_{rsts}$ ($r\ne t$), thereby facilitating fast computations of sums that enter in $V_c$ and $\lambda$'s, respectively.

\section{Program Description}
In addition to the {\tt  $<$main.c$>$} file, the package consists of eight additional files that are described below (omitting functions that are self-explanatory).

\footnotesize{

\subsection{File {\tt  $<$calculateMomentsDr.c$>$} }				
This file includes drivers that read Hamiltonians, calculate centroids, $\lambda$'s, $W(2)$'s, norms, and correlation coefficients, and write calculated quantities to output files.				
\begin{description}				
	\item {\tt 	InitModelOrb	}	 -- Performs initialization of variables for handling the orbits, Hamiltonians, and statistical measures.
	\item {\tt 	Hamiltonian	}	 -- Reads all Hamiltonians [their respective SPE's (1-body) and TBME's (2-body)].
	\item {\tt 	CalculateStatistics	}	  -- Calculates statistical moments for all Hamiltonians (including orthogonalized interactions): first, the irreducible interaction tensors (pure 0-, 1-, and 2-body) are determined; next,  second-order energy moments are calculated and stored for the base case [$n$-independent averages].  Calculates correlation coefficients and norms for $n \le 4\Omega-1$ [and $T \le \min(\frac{n}{2},2\Omega -\frac{n}{2})$ in the isospin-scalar case], according to user's choice. 
	\item {\tt 	NewModelTBMEOrthog	}	 --  Introduces a new Hamiltonian that contains the orthogonal part of a given Hamiltonian with respect to a reference one and for the new Hamiltonian, calculates quantities needed for statistical moments.  
	\item {\tt 	AllStatsOfModel	}	 --  Stores centroids and $\lambda$'s for all Hamiltonians.
	\item {\tt 	StatsOfModel	}	 --  Stores centroids and $\lambda$'s for a given Hamiltonian.    
	\item {\tt 	NullZeroLambdas	}	 -- Handles computational errors and zeroes $\lambda \le 10^{-5}$.
	\item {\tt 	AllAssignNorm	}	 -- Stores the norm for all 2-body interactions (for $n=2$ particles).
	\item {\tt 	AssignNorm	}	 -- Stores the norm for a given 2-body interaction (for $n=2$ particles).
	\item {\tt 	ClearAll	}	 -- Clears memory.
	\item {\tt 	SetClock	}	 -- Resets the clock.
	\item {\tt 	RunTime	}	 -- Calculates program run-time.
\end{description}				
\subsection{File {\tt  $<$calculateMoments.c$>$} }				
This file includes functions that calculate centroids, $\lambda$'s, $W(2)$'s, norms, and correlation coefficients.				
\begin{description}				
	\item {\tt 	TracelessSPE	}	 -- Calculates the average [Eq. (\ref{espe0})] and traceless energies  [Eq. (\ref{espe1})] for external SPE's.
	\item {\tt 	Int0b\_1b	}	 -- Calculates the centroids [Eqs. (\ref{WcS}) \& (\ref{WcTS})]  and $\lambda$'s [Eqs. (\ref{ispeS}) \& (\ref{ispeTS})] for a given  2-body interaction.
	\item {\tt 	SumV	}	 -- Calculates sums linear in $W^{JT}_{rstu}$'s [needed for computing $V_c$ and $\lambda$].
	\item {\tt 	AssnV2	}	 -- Calculates $W(2)$ [Eqs. (\ref{W2S}) \& (\ref{W2TS})] for the irreducible 2-body part of a given 2-body interaction.
	\item {\tt 	Avg2b	}	 -- Calculates Eqs. (\ref{<HH'>n2b}) \& (\ref{<HH'>nT2b}) for $\langle {\mathcal H^\dagger}_2(2) {\mathcal H'}_2(2) \rangle$.
	\item {\tt 	SumAvg2b	}	 -- Calculates the $n$-independent sum in parenthesis of Eqs. (\ref{<HH'>n2b}) \& (\ref{<HH'>nT2b}) for $\langle {\mathcal H^\dagger}_2(2) {\mathcal H'}_2(2) \rangle$.
	\item {\tt 	SumV2	}	 -- Calculates a sum quadratic in $W^{JT}_{rstu}$'s  [needed for computing second-order energy moments].
	\item {\tt 	Avg1b\_lmlm	}	 -- Calculates Eqs. (\ref{<HH'>nlm}) \& (\ref{<HH'>nTlm}) for $\langle {\mathcal H^\dagger}_2(1) {\mathcal H'}_2(1) \rangle$.
	\item {\tt 	SumAvg1b\_lmlm	}	 -- Calculates the $n$-independent sum in parenthesis of Eqs. (\ref{<HH'>nlm}) \& (\ref{<HH'>nTlm}) for $\langle {\mathcal H^\dagger}_2(1) {\mathcal H'}_2(1) \rangle$.
	\item {\tt 	Avg1b\_spelm	}	 -- Calculates Eqs. (\ref{<HH'>nspelm}) \& (\ref{<HH'>nTspelm}) for $\langle {\mathcal H^\dagger}_1(1) {\mathcal H'}_2(1) \rangle$.
	\item {\tt 	SumAvg1b\_spelm	}	 -- Calculates the $n$-independent sum in parenthesis of Eqs. (\ref{<HH'>nspelm}) \& (\ref{<HH'>nTspelm}) for $\langle {\mathcal H^\dagger}_1(1) {\mathcal H'}_2(1) \rangle$.
	\item {\tt 	Avg1b\_lmspe	}	 -- Calculates Eqs. (\ref{<HH'>nspelm}) \& (\ref{<HH'>nTspelm}) for $\langle {\mathcal H^\dagger}_2(1) {\mathcal H'}_1(1) \rangle$.
	\item {\tt 	SumAvg1b\_lmspe	}	 -- Calculates the $n$-independent sum in parenthesis of Eqs. (\ref{<HH'>nspelm}) \& (\ref{<HH'>nTspelm}) for $\langle {\mathcal H^\dagger}_2(1) {\mathcal H'}_1(1) \rangle$.
	\item {\tt 	Avg1b\_spe	}	 -- Calculates Eqs. (\ref{<HH'>nspe}) \& (\ref{<HH'>nTspe}) for $\langle {\mathcal H^\dagger}_1(1) {\mathcal H'}_1(1) \rangle$.
	\item {\tt 	SumAvg1b\_spe	}	 -- Calculates the $n$-independent sum in parenthesis of Eqs. (\ref{<HH'>nspe}) \& (\ref{<HH'>nTspe}) for $\langle {\mathcal H^\dagger}_1(1) {\mathcal H'}_1(1) \rangle$.
	\item {\tt 	AvgTot	}	 -- Calculates total averages [Eqs. (16) \& (21)] used to compute second-order energy moments for any $n$ and $T$ (calculates the variance [norm squared] in the case when $H=H'$).
	\item {\tt 	AssignAvgBaseCase	}	 -- Allocates memory and stores all $n$-independent sums in parenthesis of Eqs. (16) \& (21) for all Hamiltonians (`base case').
	\item {\tt 	SetSumAvg	}	 -- Stores all $n$-independent sum in parenthesis of Eqs. (16) \& (21) for two given Hamiltonians (`base case').
	\item {\tt 	CorrelCoeff	}	 -- Calculates correlation coefficients for given two Hamiltonians, space partitioning (scalar or isospin-scalar distribution) and number of particles.
	\item {\tt 	CreateModelTBMEOrthogonal	}	 -- Creates a new traceless 2-body Hamiltonian (with zero SPE's, $V_c$ and $\lambda$'s, by construction) that contains only the orthogonal  part of a given Hamiltonian with respect to a reference Hamiltonian according to Eq.(\ref{orthogTr0}).  
	\item {\tt 	DestroyNewModel	}	 -- Clears the last Hamiltonian that has been added during run-time to the user's list of Hamiltonians.
	\item {\tt 	StatsRefPlusOrthog	}	 -- Calculates correlation coefficients (\ref{cc_proj}), norms (\ref{norm_proj}), and projection coefficients (\ref{xLofHR}) \& (\ref{xREFofHR}) for interactions projected onto a reference and orthogonalized interactions.
	\item {\tt 	NameNewMode}  -- Provides an index for an orthogonalized Hamiltonian based on the indices of the Hamiltonian to be orthogonalized (${\rm id}$) and the reference one (${\rm id}_{\rm Ref}$). E.g. the orthogonal of ${\rm id}=23$ with respect to ${\rm id}_{\rm Ref}=22$ is indexed as $2223$.
	\item {\tt 	STmodel	}	 -- Introduces three new labels for a given Hamiltonian that correspond to a scalar distribution (all $T$), an isospin-scalar distribution for $T=0$, and an isospin-scalar distribution for $T=1$.
\end{description}				
\subsection{File {\tt  $<$tools.c$>$} \label{secTools}}				
This file contains a number of auxiliary functions, which are designed to handle s.p. basis states and TBME's. The s.p. basis states are indexed by positive integer numbers. All SPE's and TBME's are given in terms of these indices as an input. For nuclear Hamiltonians  in a HO $jj$-coupled basis, indices $k$ ($k=1,2,3,\dots$) are related to the $\eta$ (referred to as $n$ in the program) and $j$ quantum numbers of a s.p. state by $k=\frac{\eta(\eta+1)}{2} + j + \half $. For example, $k=1$ ($0s_{1/2}$),  $k=2$ ($1p_{1/2}$), $k=3$ ($1p_{3/2}$), and so on.			Two-body matrix elements $W_{ijkl}^{JT}$ are supplied only for indices,  
\begin{eqnarray*}
i \le j, \, i \le k, \,  {\rm and} 
\left\{
\begin{tabular}{l}
$ j \le l$, if $i=k$ \\
$k \le l$, if $i < k$
\end{tabular}
\right.,
\end{eqnarray*}
as all the TBME's can be then obtained using relations (\ref{conjW}). The TBME's are sorted, while being read, in ascending order in the $i-j-k-l-J$ quantum numbers and grouped according to $T$.
	
\begin{description}				
	\item {\tt 	GroupME	}	 -- Groups TBME's according to their indices (e.g., $ijij$ and $ijkj$) to reduce computational time for calculating various sums.
	\item {\tt 	allConfigurationsList	} -- Creates a model (called ``{\tt modelAllConfig}") with all possible configurations, $\{ijkl,JT\}$, that enter $V_c$ and $\lambda$'s, as well as provides checks for dimensionalities.
	\item {\tt 	FactorME	}	 --  Provides the ability to rescale TBME's by an $n$-dependent factor for a given Hamiltonian in calculating second-order energy moments. Current default: returns 1. Example given in comments is for GXPF1 \cite{HonmaOBM04} with TBME's decreasing as $[42/(40+n)]^{0.3}$.
	\item {\tt 	ScalingME	}	 --  Provides the ability to rescale TBME's  by a factor. Current default: a scaling example, 
$W^{JT}_{rstu} \rightarrow W^{JT}_{rstu}/\sqrt{(\eta_r+\eta_s)(\eta_t+\eta_u)}$, for nonzero $\eta_r,\eta_s,\eta_t,\eta_u$ and for Hamiltonians with indices between 80-89 (the rescaling is done while reading the TBME's from a file).
\end{description}				
\subsection{File {\tt  $<$propagators.c$>$} }				
This file contains propagation functions, $n$-dependent functions in Eqs. (\ref{<HH'>nspe}-\ref{<HH'>n2b}) for the scalar case and Eq. (\ref{propagators}) for the isospin-scalar case, used to calculate second-order energy moments for a many-particle system (for any number of particles $n$ and isospin $T$).				

\subsection{I/O Files {\tt  $<$readFile.c$>$} and {\tt  $<$writeFile.c$>$}}				
The functionalities in  {\tt  $<$readFile.c$>$}  include reading the user input menu as well as reading and storing (in arrays or linked lists) SPE's and TBME's of a set of Hamiltonians specified by the user.  				
				
The {\tt  $<$writeFile.c$>$} consists of functions for writing the output. In the output files, the number of particles is denoted as $m$ and the isospin values are given as twice the value, $2T$.  	
\begin{description}				
	\item {\tt 	PrintListModel	}	 -- Prints TBME's to a file for all $T$ for a given Hamiltonian.
	\item {\tt 	PrintListModelT	}	 -- Prints TBME's to a file for a specific $T$  for a given Hamiltonian.
	\item {\tt 	PrintLambda	}	 -- Prints $\lambda$'s for a given Hamiltonian and space partitioning (and isospin).
	\item {\tt 	PrintLambdaH	}	 -- Prints $\lambda$'s for all Hamiltonians and a given space partitioning.
	\item {\tt 	PrintStatsOfModel	}	 --  Prints centroids, $\lambda$'s, and norms of a Hamiltonian.
	\item {\tt 	PrintStatsOfAllModel	}	 -- Prints centroids, $\lambda$'s, and norms for all Hamiltonians.
	\item {\tt 	PrintNorm	}	 -- Prints norms for all Hamiltonians given a space partitioning for all $n$ (and $T$).
	\item {\tt 	PrintNormPairsModel	}	 -- Prints norms for all Hamiltonians for a given $n$ (and $T$).
	\item {\tt 	PrintCorrelCoef	}	 -- Prints correlation coefficients for all Hamiltonians given a space partitioning for all $n$ (and $T$).
	\item {\tt 	PrintCCPairsModel	}	 -- Prints  correlation coefficients for all pairs of Hamiltonians for a given $n$ (and $T$), e.g., $(H_1,H_2)$ $(H_1,H_3)$ $(H_2,H_3)$.
	\item {\tt 	PrintCCPairsOneModel	}	 -- Prints  correlation coefficients for a given Hamiltonian for a given $n$ (and $T$), e.g., for $H_1$, $(H_1,H_2)$ $(H_1,H_3)$.
	\item {\tt 	PrintRefOrthogQnty	}	-- Prints correlation coefficients (\ref{cc_proj}), norms (\ref{norm_proj}), and projection coefficients (\ref{xLofHR}) \& (\ref{xREFofHR}) for interactions projected onto a reference and orthogonalized interactions.
	\item {\tt 	PrintCorrelCoefEachUnitaryRk	}	 -- Prints correlation coefficients for irreducible 1-body and 2-body interaction tensors.
	\item {\tt 	PrintList2mdls	}	 -- (additional) Prints TBME's of two models for the same set of quantum numbers [zero TBME's are included if missing in one model, but existing in the other; also, every TBME of $J$ and $T$ is printed $(2J+1)(2T+1)$ times].
	\item {\tt 	TestOrthogMdls	}	 -- Provides a test if an orthogonalized Hamiltonian is orthogonal to its reference Hamiltonian.
	\item {\tt   PrintInitNorm, PrintInitCCheader, PrintInitCC(OneModel), PrintInitRefOrthogQnty	}	 -- Write various headers that provide labels for each type of calculations in the output file.
\end{description}				
\subsection{Files {\tt  $<$arrayFns.c$>$} and {\tt  $<$linkList.c$>$} }				
These files contain conventional array and linked-list manipulations.				
\subsection{Data-type definitions and variables}				
\begin{description}				
	\item {\tt 	caseST	}	-- Data type for distributions: {\tt kScalar} (or 0) if scalar distribution; {\tt kTScalar} (or 1) if isospin-scalar distribution.
	\item {\tt 	orbType	}	-- Data type for indices: {\tt jqnOrb} refers to the $2j$ angular momentum of a s.p. state; {\tt indexOrb} refers to the index $k=1,2,3,...$ of a s.p. state.
	\item {\tt 	intPart	}	-- Data type for irreducible tensors: {\tt hAll} specifies the entire 1-body+2-body Hamiltonian (all irreducible tensors), {\tt hLm2b} refers to ${\mathcal H}_2(1)+{\mathcal H}_2(2)$, {\tt h2b} refers to ${\mathcal H}_2(2)$, and {\tt hLm} refers to ${\mathcal H}_2(1)$.
	\item {\tt 	meType 	}	-- Data type for TBME's $W_{ijkl}^{JT}$: 
	{\tt 	kiOrb, kjOrb, kkOrb, klOrb	}	[s.p. indices $i$, $j$, $k$, $l$];
	{\tt 	kJ } and {\tt 	kT	}	[$2J$ and $2T$];
	{\tt 	dimJ } and {\tt 	 dimJT	}	[$2J+1$ and  $(2J+1)(2T+1)$];
	{\tt 	vInt } and {\tt 	 vInt2	}	[$W^{JT}_{ijkl}$ and $W^{JT}_{ijkl}(2)$];
	{\tt 	numericME} and {\tt 	numericME2	}	[true if  $W^{JT}_{ijkl}$ and $W^{JT}_{ijkl}(2)$ assigned].
	\item {\tt 	statType	}	-- Data type for key characteristics of a  Hamiltonian, namely, $V_c$, $\lambda$'s and norms.
	\item {\tt 	avgBCType	}	-- Data type for base-case averages. 
	\item {\tt 	arrayModels, arrayOrbits, arrayDimOrb, arrayRadDeg, numRadDeg, arraySPE}	-- Arrays to store Hamiltonian id's, orbits, dimensionalities, radial-degenerate orbits, number of radial-degenerate orbits for a given orbit, and SPE's, respectively.
	\item {\tt 	arrayStat, arrayAvgBaseCase	}	-- Arrays to store  $V_c$, $\lambda$'s, and norms of Hamiltonians ({\tt arrayStat}), as well as  base-case averages ({\tt 	arrayAvgBaseCase}).
	\item {\tt 	head	}	 -- Array of linked-list heads for TBME's of all Hamiltonians.
	\item {\tt 	kModels, kOrbits, kShells, kOrthogMdls	}	-- Provides the total number of Hamiltonians, orbits, major HO shells, and orthogonalized Hamiltonians, respectively.
	\item {\tt 	numModels, numOrbits, orthogInModels	}	-- Provides the maximum allowed number of Hamiltonians [including orthogonalized Hamiltonians and {\tt modelAllConfig} that is automatically created], orbits, and orthogonalized Hamiltonians, respectively.
\end{description}				
			
} 

\section*{Acknowledgments}

This work was supported by the U.S.  National Science Foundation (OCI-0904874), 
the U.S. Department of Energy (DE-SC0005248), and the 
Southeastern Universities Research Association.

\section*{Appendix}

\noindent
{\tiny 
{\bf \footnotesize Sample input files for Hamiltonians}\\ \\
Part of a sample input file for SPE's ($\varepsilon_{k}$ is specified as {\tt $<$ $k$  $\varepsilon(MeV)$ $>$}, with indices $k=1,2,3,\dots$):
\begin{verbatim}
1 1.0532
2 1.8217
3 2.0244
...
\end{verbatim}
\noindent
Part of a sample input file for TBME's ($W^{JT}_{ijkl}$ is specified as {\tt $<$ $i$ $j$ $k$ $l$ $2J$ $2T$ $W(MeV)$  $>$}, with indices $i,j,k,l=1,2,3,\dots$):
\begin{verbatim}
1	1	1	1	2	0	0.803376
1	1	1	4	2	0	-7.684559
1	1	2	2	2	0	-1.161099
1	1	2	7	2	0	-1.377213
1	2	1	2	2	0	5.997909
2	9	7	8	4	2	-5.486247
...
\end{verbatim}
For further details, see Sec. \ref{secTools}.\\ \\
\noindent
{\bf \footnotesize Program input file}\\
Test input (file) for the output shown below:
\begin{verbatim}
sp4_6int21_22_23
3               /* NO. of orbits
4 5 6           /* orbits
3               /* NO. of interactions
21 22 23        /* interactions (TBME)
1               /* 0 sets all external s.p.energies=0 for all interactions; 1 reads from files
2               /* output: (1) CC; (2) CC+norms; (3) 2 + additional CC/norms, (4) 1+test
2               /* NO. of interactions to be orthogonalized
22 23           /* Href & Interaction to be orthogonalized to Href
23 21
\end{verbatim}
\noindent
Sample input (file) for TBME rescaling:
\begin{verbatim}
sp4_10int81_22_23
7                       /* NO. of orbits
4 5 6 7 8 9 10          /* orbits
3                       /* NO. of interactions MESCALED
81 22 23                /* interactions (TBME)
0                       /* 0 sets all external s.p.energies=0 for all interactions; 1 reads from files
3                       /* output: (1) CC; (2) CC+norms; (3) 2 + additional CC/norms, (4) 1+test
1                       /* NO. of interactions to be orthogonalized
22 23                   /* Href & Interaction to be orthogonalized to Href
\end{verbatim}
\noindent
{\bf  \footnotesize  User input menu:}
\begin{verbatim}
Enter the name to label output files

		USER INPUT MENU
		---------------

********************** ORBITS **************************
MAXIMUM ALLOWED ORBITS: 45
Enter -1 to terminate program

Sample Input :-
NO. OF ORBITS => 3
ORBITS => 1 2 3
********************************************************
Enter number of orbits:
Enter orbits in ASCENDING order:


ORBITS READ: 4 5 6 


******************* INTERACTIONS ***********************
Interactions are represented by numerical indices corresponding
to their respective input files of two-body matrix elements, TBMEs
(e.g. '8' --> TBME8.dat, '13' --> TBME13.dat)

Interactions with indices between 80-89 can be rescaled
(refer to ScalingME in tools.c).
For rescaling: add a comment containing MESCALED after entering number of interactions.

MAX. ALLOWED INTERACTIONS: 20
Enter -1 to terminate program

Sample Input :-
NO. OF INTERACTIONS => 2
INTERACTIONS => 21 23
********************************************************
Enter number of interactions (followed by MESCALED for rescaling):
Enter interactions:


INTERACTIONS READ: 21 22 23 


********** SINGLE PARTICLE ENERGIES (SPE's) ************
Single particle energies will be read from files corresponding to
the interaction numerical indices
(e.g. '21'---> SPE21.dat, '22' ---> SPE22.dat)

0. No external SPE's (for all interactions)
1. Reads SPE's from files (for all interactions)
Enter -1 to terminate program
********************************************************
Enter (1/0/-1):


SPE's INCLUDED


******************** OUTPUT MENU ***********************
1. Correlation Coefficients (CC's)
2. Correlation Coefficients (CC's) and Norms
3. 2 + CC's for Irreducible (1-body and 2-body) Parts of Interactions
4. 1 + Tests for orthogonality 
********************************************************
Enter (1/2/3/4):


CHOICE FOR OUTPUT: 2


********** INTERACTIONS TO BE ORTHOGONALIZED ***********
Interactions must be chosen from Interactions specified above.

MAX. ALLOWED PAIRS: 3
Enter -1 to terminate program

Sample Input :-
NO. OF INTERACTIONS TO BE ORTHOGONALIZED => 2
REF. AND ORTHOG. MODELS #1 => 22 23
REF. AND ORTHOG. MODELS #2 => 23 21
********************************************************
Enter number of interactions to be orthogonalized:
Enter pairs: Reference Interaction Href and Interaction to be orthogonalized to Href 
Enter pair #1:
Enter pair #2:


INTERACTIONS TO BE ORTHOGONALIZED: 23 (to 22)   21 (to 23)   

		---------------

::InitModelOrb:: Finish reading input ...
::Hamiltonian:: Reading SPE/TBME...
::CalculateStatistics:: Calculating moments...
::CalculateStatistics:: Scalar distribution...Finish calculating...
::CalculateStatistics:: Isospin-scalar distribution...Finish calculating...
::ClearAll:: End.
\end{verbatim}
\noindent
{\bf \footnotesize  Test output}\\
File {\tt $<$out\_sp4\_6int21\_22\_23.dat$>$}, shown only for Scalar distribution (format: semicolon-delimited):
\begin{verbatim}
***************************************************************************
                        *Scalar distribution*                              
***************************************************************************
Hamiltonian;         21;         22;         23;       2223;       2321;
Vc;            -2.81556;    0.11436;   -1.66531;   -0.00000;   -0.00000;
norm;           9.82673;    9.43303;   10.35063;   10.18719;    9.81476;
lambda:
 4  4;         14.51213;   -0.91548;   12.97585;    0.00000;    0.00000;
 5  5;          5.99441;  -24.58659;   -4.52222;    0.00000;    0.00000;
 6  6;         -8.83365;   16.69622;   -1.31047;    0.00000;    0.00000;

Correlation Coefficients (H1,H2), Projection Norm |Hproj=Href+c*Hperp| and Projection Coefficients cH_H1 of H along H1:
  m;     (21,22);   (21,2223);(21,22+2223);  |21proj|;  c21_22;c21_2223;     (21,23);   (21,2321);(21,23+2321);  |21proj|;  c21_23;c21_2321;
  2;     0.07529;     -0.0278;     0.08026; 1.006e+01; 0.07843; -0.0268;     -0.0160;     0.99878;     0.99891; 6.464e+02; -0.0152; 1.00000;
  3;     0.06932;     -0.0278;     0.07469; 1.693e+01; 0.07154; -0.0268;     -0.0149;     0.99977;     0.99988; 1.147e+03; -0.0141; 1.00000;
  4;     0.06269;     -0.0278;     0.06858; 2.332e+01; 0.06419; -0.0268;     -0.0137;     0.99882;     0.99892; 1.680e+03; -0.0130; 1.00000;
  5;     0.05546;     -0.0278;     0.06202; 2.949e+01; 0.05632; -0.0268;     -0.0123;     0.99725;     0.99733; 2.282e+03; -0.0117; 1.00000;
  6;     0.04755;     -0.0277;     0.05504; 3.573e+01; 0.04787; -0.0268;     -0.0108;     0.99528;     0.99533; 3.010e+03; -0.0103; 1.00000;
  7;     0.03890;     -0.0276;     0.04772; 4.275e+01; 0.03879; -0.0268;     -0.0091;     0.99295;     0.99299; 3.974e+03; -0.0087; 1.00000;
  8;     0.02938;     -0.0276;     0.04028; 5.252e+01; 0.02900; -0.0268;     -0.0072;     0.99026;     0.99029; 5.437e+03; -0.0069; 1.00000;
  9;     0.01886;     -0.0275;     0.03333; 7.284e+01; 0.01840; -0.0268;     -0.0051;     0.98717;     0.98718; 8.203e+03; -0.0048; 1.00000;
 10;     0.00717;     -0.0274;     0.02830; 1.722e+02; 0.00691; -0.0268;     -0.0026;     0.98360;     0.98361; 1.656e+04; -0.0025; 1.00000;
 11;     -0.0059;     -0.0273;     0.02790; 2.146e+02; -0.0056; -0.0268;     0.00023;     0.97946;     0.97946; 1.903e+05; 0.00022; 1.00000;
 12;     -0.0206;     -0.0271;     0.03408; 7.724e+01; -0.0193; -0.0268;     0.00355;     0.97462;     0.97463; 1.251e+04; 0.00341; 1.00000;
 13;     -0.0374;     -0.0270;     0.04607; 5.861e+01; -0.0343; -0.0268;     0.00746;     0.96889;     0.96892; 5.883e+03; 0.00719; 1.00000;
 14;     -0.0565;     -0.0268;     0.06255; 5.282e+01; -0.0509; -0.0268;     0.01216;     0.96200;     0.96208; 3.514e+03; 0.01176; 1.00000;
 15;     -0.0788;     -0.0265;     0.08312; 4.999e+01; -0.0692; -0.0268;     0.01789;     0.95359;     0.95375; 2.284e+03; 0.01739; 1.00000;
 16;     -0.1049;     -0.0263;     0.10815; 4.787e+01; -0.0897; -0.0268;     0.02506;     0.94307;     0.94341; 1.529e+03; 0.02449; 1.00000;
 17;     -0.1362;     -0.0259;     0.13863; 4.571e+01; -0.1126; -0.0268;     0.03426;     0.92957;     0.93020; 1.023e+03; 0.03374; 1.00000;
 18;     -0.1744;     -0.0254;     0.17624; 4.319e+01; -0.1385; -0.0268;     0.04651;     0.91160;     0.91279; 6.693e+02; 0.04628; 1.00000;
 19;     -0.2224;     -0.0247;     0.22377; 4.014e+01; -0.1679; -0.0268;     0.06365;     0.88649;     0.88877; 4.168e+02; 0.06423; 1.00000;
 20;     -0.2851;     -0.0236;     0.28605; 3.642e+01; -0.2017; -0.0268;     0.08933;     0.84888;     0.85357; 2.381e+02; 0.09209; 1.00000;
 21;     -0.3716;     -0.0219;     0.37225; 3.185e+01; -0.2410; -0.0268;     0.13215;     0.78609;     0.79712; 1.164e+02; 0.14115; 1.00000;
 22;     -0.5025;     -0.0183;     0.50287; 2.613e+01; -0.2870; -0.0268;     0.21849;     0.65796;     0.69328; 4.129e+01; 0.25049; 1.00000;
 23;     -0.7390;     0.00000;     0.73901; 1.847e+01; -0.3419; 0.00000;     0.49577;     0.00000;     0.49577; 5.978e+00; 0.70846; 0.00000;
Norms |H|:
  m;    |  21|;    |  22|;    |  23|;    |2223|;    |2321|;
  2; 9.827e+00; 9.433e+00; 1.035e+01; 1.019e+01; 9.815e+00;
  3; 1.621e+01; 1.571e+01; 1.711e+01; 1.682e+01; 1.621e+01;
  4; 2.183e+01; 2.132e+01; 2.303e+01; 2.263e+01; 2.180e+01;
  5; 2.678e+01; 2.637e+01; 2.823e+01; 2.772e+01; 2.670e+01;
  6; 3.108e+01; 3.088e+01; 3.274e+01; 3.211e+01; 3.094e+01;
  7; 3.476e+01; 3.485e+01; 3.656e+01; 3.582e+01; 3.451e+01;
  8; 3.780e+01; 3.830e+01; 3.970e+01; 3.885e+01; 3.743e+01;
  9; 4.022e+01; 4.122e+01; 4.217e+01; 4.121e+01; 3.970e+01;
 10; 4.201e+01; 4.360e+01; 4.397e+01; 4.289e+01; 4.132e+01;
 11; 4.318e+01; 4.545e+01; 4.509e+01; 4.390e+01; 4.230e+01;
 12; 4.373e+01; 4.676e+01; 4.554e+01; 4.424e+01; 4.262e+01;
 13; 4.365e+01; 4.752e+01; 4.531e+01; 4.390e+01; 4.230e+01;
 14; 4.296e+01; 4.773e+01; 4.441e+01; 4.289e+01; 4.132e+01;
 15; 4.164e+01; 4.738e+01; 4.285e+01; 4.121e+01; 3.970e+01;
 16; 3.969e+01; 4.644e+01; 4.060e+01; 3.885e+01; 3.743e+01;
 17; 3.713e+01; 4.491e+01; 3.769e+01; 3.582e+01; 3.451e+01;
 18; 3.394e+01; 4.274e+01; 3.411e+01; 3.211e+01; 3.094e+01;
 19; 3.012e+01; 3.990e+01; 2.985e+01; 2.772e+01; 2.670e+01;
 20; 2.568e+01; 3.629e+01; 2.491e+01; 2.263e+01; 2.180e+01;
 21; 2.062e+01; 3.180e+01; 1.931e+01; 1.682e+01; 1.621e+01;
 22; 1.492e+01; 2.612e+01; 1.301e+01; 1.019e+01; 9.815e+00;
 23; 8.543e+00; 1.847e+01; 5.978e+00; 0.000e+00; 0.000e+00;
 \end{verbatim}
\noindent
} 
\noindent
This output file provides the centroid (listed as ``Vc"), the norm, as well as the density-dependent 1-body part (listed as ``lambda") for each Hamiltonian, together with various second-order moments listed for increasing number of particles (shown as ``m") for the scalar distribution, and also for twice the isospin value (shown as ``2T") for the isospin-scalar distribution. In a complete space (e.g., $24$ particles for the output shown above, m$=4\Omega$), moments are zero by definition, Eqs. (16) and (\ref{propagators}), and the corresponding output is omitted.

Additional information, including dimensionality checks, set of input parameters, run-time, and messages on rescaling and orthogonalization, are written into a corresponding log-file \tiny{ {\tt $<$o\_sp4\_6int21\_22\_23.log$>$}}.

\end{document}